\documentclass[aps,prl,twocolumn,10pt,amsmath,amssymb,bibnotes,superscriptaddress,longbibliography]{revtex4-1}

\usepackage{graphics,bm,amsmath,amssymb,natbib,url,epsfig,psfrag, color}

\newcommand{\be}{\begin{equation}}
\newcommand{\ee}{\end{equation}}
\newcommand{\bea}{\begin{eqnarray}}
\newcommand{\eea}{\end{eqnarray}}

\begin{document}
	
	\title{Probing  single-electron scattering through a non-Fermi liquid charge-Kondo device}

	\author{Eran Sela}
	\affiliation{Raymond and Beverly Sackler School of Physics and Astronomy, Tel-Aviv University, IL-69978 Tel Aviv, Israel}
	\author{David Goldhaber-Gordon}
	\affiliation{Department of Physics, Stanford University, Stanford, CA 94305}
	\affiliation{
		Stanford Institute for Materials and Energy Sciences,
		SLAC National Accelerator Laboratory, Menlo Park, CA 94025}
	\author{A. Anthore}
	\affiliation{Universit\'e Paris-Saclay, CNRS, Centre de Nanosciences et de Nanotechnologies (C2N), 91120 Palaiseau, France}
	\affiliation{Université Paris Cité, CNRS, Centre de Nanosciences et de Nanotechnologies, F-91120, Palaiseau, France}
	\author{F. Pierre}
	\affiliation{Universit\'e Paris-Saclay, CNRS, Centre de Nanosciences et de Nanotechnologies (C2N), 91120 Palaiseau, France}
	\author{Yuval Oreg}
	\affiliation{Department of Condensed Matter Physics, Weizmann Institute of Science, Rehovot, 76100, Israel}
	\date{\today}
	
	\begin{abstract}
		Among the exotic and yet unobserved features of multi-channel Kondo impurity models is their sub-unitary single electron scattering. In the two-channel Kondo model, for example, an incoming electron is fully scattered into a many-body excitation such that the single particle Green function vanishes. Here we propose to directly observe these features in a charge-Kondo device encapsulated in a Mach-Zehnder interferometer - within a device already studied in Ref.~\onlinecite{duprez2019transmitting}. We provide detailed predictions for the visibility and phase of the Aharonov-Bohm oscillations depending on the number of coupled channels and the asymmetry of their couplings. 
	\end{abstract}
	
	\maketitle

	\paragraph*{Introduction.} The conventional Kondo effect is described by a local Fermi liquid theory
	~\cite{nozieres1974fermi}. One of its manifestations is the scattering phase shift resulting in a Kondo resonance forming at the Fermi level. The multi-channel Kondo (MCK) model, however, is described by a non-Fermi liquid (NFL) theory~\cite{nozieres1980kondo}, and even the scattering of an electron incident at the Fermi level is inelastic~\cite{borda2007theory}: it cannot be described by a single particle scattering phase shift. For the specific and most dramatic case of two channels, the incoming electron scatters purely into a many-body excitation~\cite{affleck1993exact,emery1992mapping} with no elastic scattering, which is often referred to as the unitarity paradox~\cite{maldacena1997majorana}. For three or more channels there is a finite (yet non-unitary) elastic scattering probability which tends to unity in the limit of a large number of channels~\cite{affleck1993exact}. Though quantum dot (QD) experiments have proved successful in verifying many predictions on electronic transport through Kondo impurities, a direct observation of the single electron scattering amplitude in NFL states has remained elusive. 
	
	An experimental proposal~\cite{carmi2012transmission} to measure the NFL single electron scattering consists of embedding a spin-Kondo QD~\cite{oreg2003two,potok2007observation} in one arm of an Aharonov-Bohm (AB) interferometer. In this geometry, half of the conductance is expected to be incoherent while the other half is coherent with a definite phase shift~\cite{carmi2012transmission} since only one linear combination of the source and drain leads couples to the QD while the other remains free, described by a FL theory. Realizing this setup with the necessary control of all parameters has been attempted by one of the present authors, but it has proved  challenging.
	
\begin{figure}[]
		\includegraphics[width=1\columnwidth]{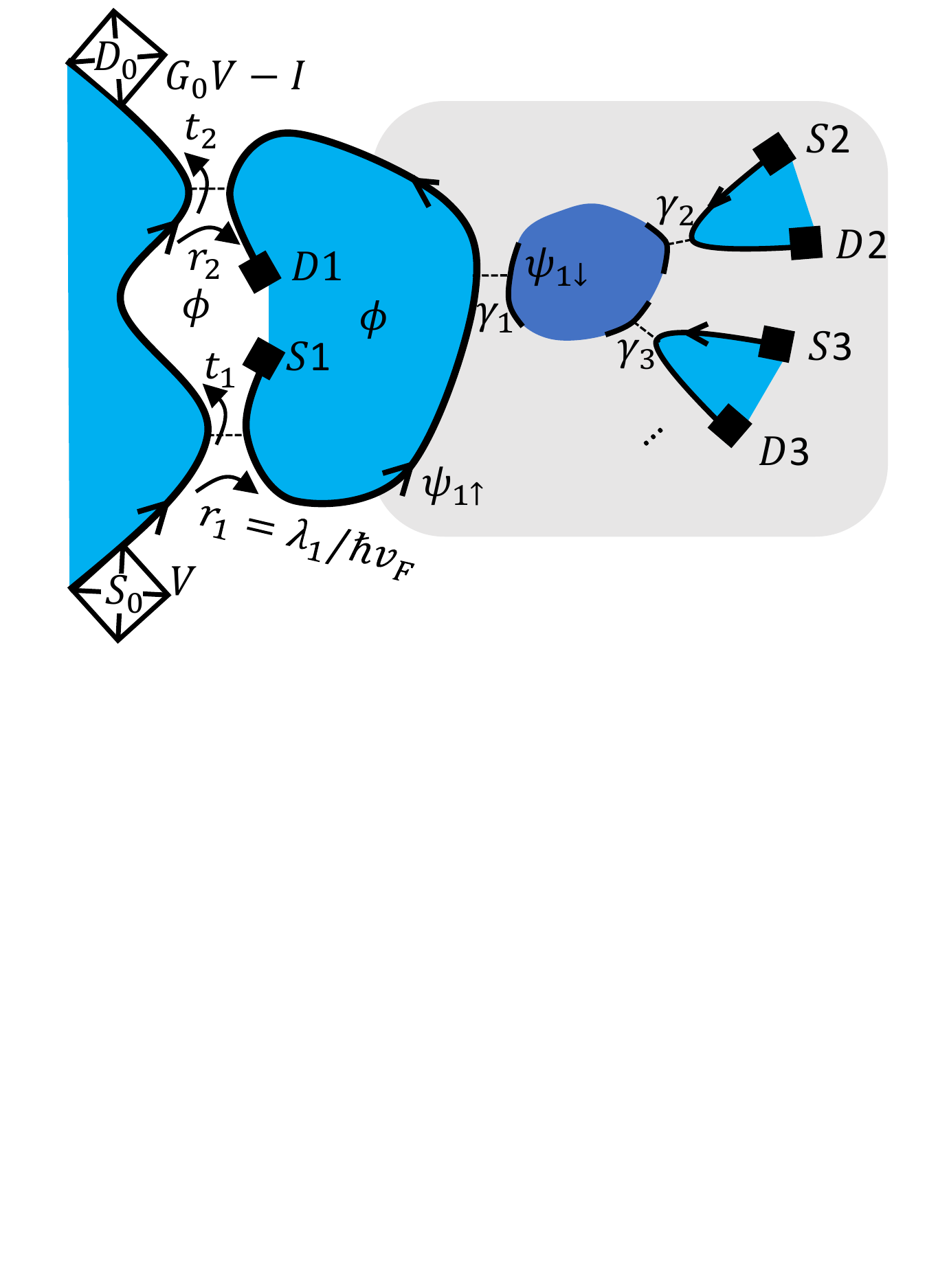}
		\vspace{-6cm}
		\caption{Schematics of the device of Ref.~\onlinecite{duprez2019transmitting} based on chiral edge states. The blue areas are in the $\nu=1$ quantum Hall state, with black boundaries denoting the propagating edge mode.	The grey box including the dark blue confined dot highlights the Multi-channel Kondo system. The Mach Zehnder interferometer contains two QPCs  with  reflection  amplitudes $r_1$ and $r_2$. In the text we analyzed the differential conductance $dI/dV$.}\label{fig:1}
	\end{figure}

	In this paper we propose a complementary approach for observing the single electron scattering amplitude and phase in the Kondo effect, using a multi-channel charge-Kondo system~\cite{iftikhar2015two,iftikhar2018tunable} based on quantum Hall edge states, embedded in a Mach–Zehnder interferometer. 
	This experimental configuration was recently designed by Duprez \emph{et. al.}~\cite{duprez2019transmitting}, see Fig.~\ref{fig:1}. 
	Here, we argue that such a device is capable of directly identifying so-far-elusive information about the NFL nature of the MCK effect.

	\paragraph*{Model.}
	The right hand side of the device in Fig.~\ref{fig:1}, demarcated by a gray box, is the charge-multichannel Kondo system. Recent experiments~\cite{iftikhar2015two,iftikhar2018tunable} matched detailed conformal field theory predictions for the conductance of two and three channels: $G_{2CK}=\frac{e^2}{2h}$ and $G_{3CK}=\frac{4e^2}{3h}\sin^2 \frac{\pi}{5}$, respectively, at the intermediate-coupling fixed points, as well as renormalization group flows toward the single channel fixed point as asymmetry between the channels becomes apparent. 
	
	In this charge-based realization of the Kondo effect~\cite{Matveev1991,matveev1995coulomb,furusaki1995theory}, two charge states of the QD play the role of the impurity spin: 
	the $N+1$ charge state is mapped to ``spin up"- and the $N$ charge state to ``spin down". Consider weakly transmitting quantum point contacts (QPC) coupling the dot to the leads. Each reflection process in the QPC (i.e. $\gamma_i$ in Fig.~\ref{fig:1}) translates into an impurity ``spin flip". To complete the analogy to spin, label the annihilation operators of spinless electrons in each lead ($j=1,2,3$ in the figure) by $\psi_{j \uparrow}$, and spinless electrons inside the QD near each QPC by $\psi_{j \downarrow}$ as marked in Fig.~\ref{fig:1}. As the Kondo effect requires a continuum in both spin flavors, the electronic level spacing in the QD must be negligible. In practice, incorporating a metal (with its high density of states) into the QD has allowed achieving this criterion without making the charging energy unacceptably small.

	Thus, the free part of the Hamiltonian $H=H_0+H_K$, describes $2_{QD / lead}  \times 3_{QPCs}$ chiral channels, $H_0=\sum_{\alpha=\uparrow,\downarrow,j=1,2,3} \int dx \psi_{j \alpha}^\dagger(x) i v_F \partial_x \psi_{j \alpha}(x)$. There is no elastic transport between different QPCs due to the large ratio between the temperature and the level spacing in the metallic QD~\cite{Matveev1991,matveev1995coulomb,furusaki1995theory}. The Kondo interaction simply describes tunneling in and out of the QD~\cite{mitchell2016universality},
	\be
	H_K=\sum_j \gamma_j \psi^\dagger_{j \downarrow}(0) \psi_{j \uparrow}(0) S^+ + h.c.+\Delta E S^z,
	\ee
	with amplitudes $\gamma_j$.
	Here $S^+=|N+1\rangle \langle N|$,$S^z=( |N+1 \rangle \langle N+1 |-|N \rangle \langle N |)/2$, and $\Delta E \propto V_g-V_g^0$ describes a gate-voltage-dependent energy splitting between the two charge states. 
	
	The MZ interferometer in Fig.~\ref{fig:1} consists of two arms denoted 0 and 1. The zeroth (``reference") arm in the left side in Fig.~\ref{fig:1} is described by a chiral fermion $H_{0,ref}=\int dx  \psi_{0}^\dagger(x) i v_F \partial_x \psi_{0}(x)$. Arm  1 is described by $\psi_{1 \uparrow}$. 
	The tunneling Hamiltonian of the MZ interferometer is 
	\bea
	H_{tun}=\lambda_1 \psi^\dagger_0 (-L) \psi_{1 \uparrow } (-L)+\lambda_2 e^{i \phi} \psi^\dagger_0 (L) \psi_{1 \uparrow } (L
	)+H.c.,
	\eea
	where $\phi$ 
	is the AB flux in units of $h/e$.

	\paragraph*{Interference current.}
	Useful information about the MCK state can be extracted from the current within second order perturbation theory in the tunneling amplitudes $\lambda_{1,2}$. To leading order these are given in terms of the reflection amplitudes at the MZ QPCs, $r_i=\lambda_i/(\hbar v_F)$~\cite{SM} ($i=1,2$). In the absence of the QD ($\gamma_1=0$), the conductance between arm 0 and arm 1
	\be
	\label{eq:simple}
	\left.\frac{dI}{dV}\right|_{\gamma_1=0}=\frac{e^2}{h} \left|r_1+r_2 e^{i \phi}\right|^2.
	\ee
	The current can be separated as $I=I_0+I_{\Phi}$, where $I_0 \propto |r_1|^2+|r_2|^2$ does not depend on the AB flux, and $I_{\Phi}$ is the interference term, which in the absence of the QD ($\gamma_1=0$) is $dI_{\Phi}/dV=\frac{e^2}{h} ( r_1 r_2^* e^{-i \phi}+h.c.)$. 
	
	Calculating the current using the Kubo formula to second order in the reflection, one finds
	~\cite{law2006electronic,duprez2019transmitting} 
	\bea
	\label{eq:gf}
	I_\Phi&=& r_1 r_2^* e^{-i \phi} \int_{-\infty}^\infty dt e^{i Vt} \times  \\
	&&(\langle \psi^\dagger_0(-L) \psi_0(L,t) \rangle  \langle \psi_1(-L) \psi^\dagger_1(L,t) \rangle \nonumber \\
	&-&\langle \psi_0(L,t) \psi^\dagger_0(-L) \rangle \langle \psi^\dagger_1(L,t) \psi_1(-L) \rangle) + h.c.\nonumber
	\eea
	We see that the interference term probes the electron propagator  between $x=-L$ and $x=L$, which for arm 1 is sensitive to the scattering amplitude of the QD.

	The interference term $r_1 r_2^* e^{-i \phi}$ in Eq.~(\ref{eq:simple}) corresponds to an electron passing uninterrupted through channel 1, unaffected by the QD. This is obtained by substituting the  free Green function denoted $G_0$ for both arms~\cite{SM}. In this case the variation of the conductance due to the AB oscillations is given by $4|r_1 r_2|$. The smallness of $r_1$ and $r_2$ guarantees that the interference term is due to a single electron being injected into arm 1. Practically, it is sufficient to demand that only $|r_1| \ll 1$ whereas $|r_2|$ can be of order unity.
	
	All nonzero orders in the $\gamma_j$'s 
	are encoded by the exact $\mathcal{T}$-matrix~\cite{pustilnik2004kondo,pustilnik2001kondo}, $G(\omega,-L,L)=G_0(\omega,-L,L)+G_0(\omega,-L,0) \mathcal{T}(\omega) G_0(\omega,0,L)$. Substituting into Eq.~(\ref{eq:gf}) yields at zero temperature~\cite{sela2009nonequilibrium,SM} 
	\be
	G_\Phi \equiv \frac{h}{e^2}\frac{dI_{\Phi}}{dV}
	=r_1 r_2^* e^{-i \phi} \mathcal{S}(eV)+h.c.,
	\ee
	where
	\be
	\mathcal{S}(\omega)=1-2\pi i \nu \mathcal{T}
	(\omega).
	\ee
	As a matrix, $\mathcal{S}$ is diagonal in both on-site pseudospin and channel index, and we refer throughout to the matrix elements $\mathcal{S}_{1 \uparrow, 1 \uparrow}$  as probed by the MZ interferometer (and similarly for $\mathcal{T}$).
	Since this result is perturbative in the tunneling between arms 0 and 1, the voltage $V$ sets the value of the frequency of the $\mathcal{T}$ matrix computed at equilibrium.
	
	From the $\mathcal{S}$-matrix, we see that the visibility, i.e. the difference in conductance in units of $e^2/h$ between maximum and minimum, is given by $4|r_1 r_2| |\mathcal{S}|$, and the phase of the AB oscillations is given by $\arg \mathcal{S}$,
	\be
	\label{eq:genericform}
	G_\Phi(V,T)=2 |r_1 r_2| |\mathcal{S}(eV)| \cos(\arg (r_1 r_2^*)+\arg (\mathcal{S}(eV)) -\phi).
	\ee
	We refer to $|\mathcal{S}|$ as the \emph{relative visibility} with respect to the trivial case in Eq.~(\ref{eq:simple}). 
	

	\begin{figure*}[]
		\includegraphics[width=2\columnwidth]{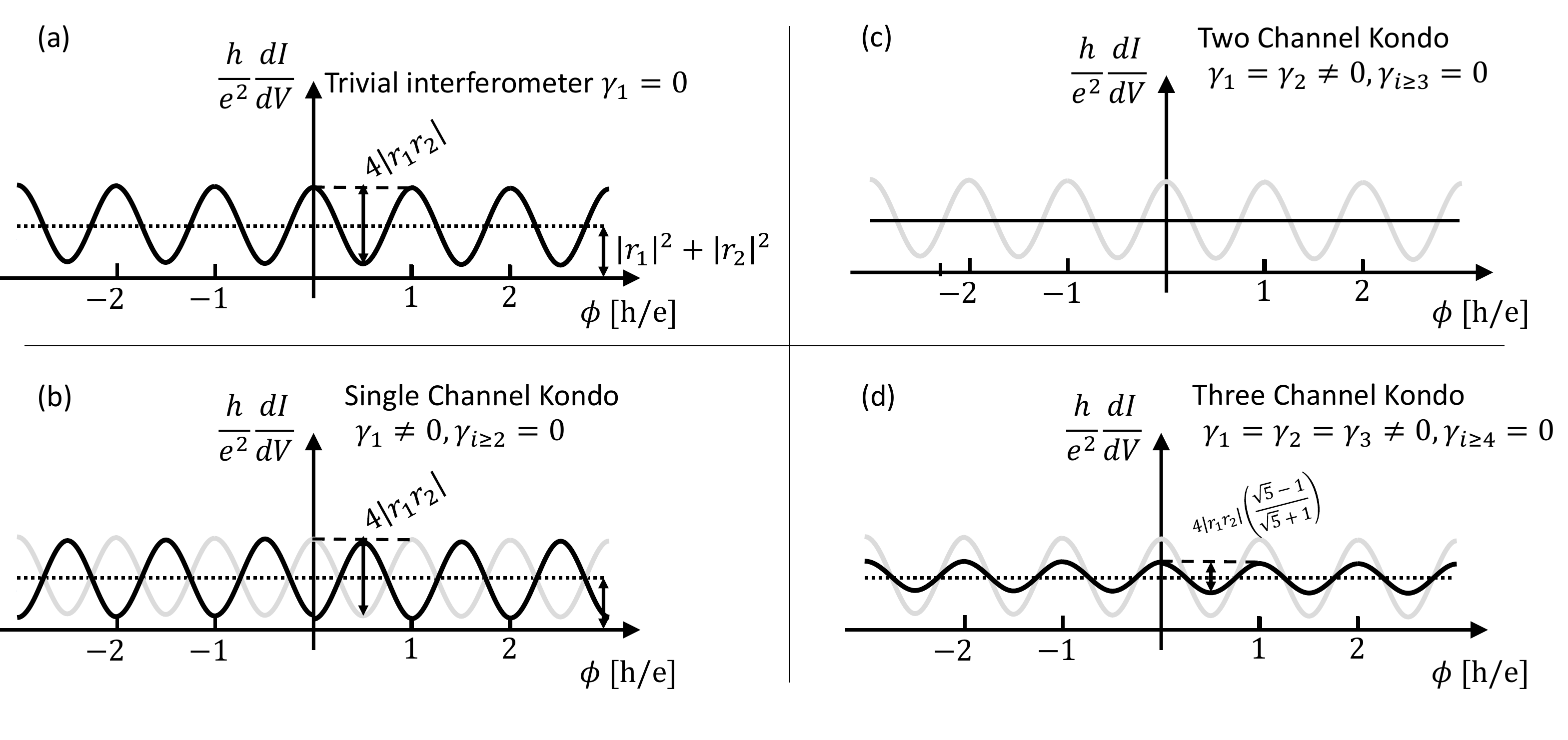}
		\caption{AB oscillation patterns for (a) the trivial case with a detached  QD, and for (b) 1CK, (c) 2CK and (d) 3CK configurations. Both the amplitude of the phase of the single electron scattering amplitude $\mathcal{S}$ in the $k-$channel Kondo state can be read by comparison with the trivial reference case of (a).  }\label{fig:ABosc}
	\end{figure*}
	
	Consider first the case where $k$ QPCs are equally coupled ($\gamma_1 = \gamma_2 = \dots = \gamma_k$), and where the gate voltage of the QD is tuned to an exact degeneracy between the two charge states ($\Delta E=0$). In this case we employ the Affleck and Ludwig 
	result 
	for the  $\mathcal{S}$-matrix at the $k$-channel Kondo fixed point~\cite{affleck1993exact},
	\be
	\label{ALresult}
	\mathcal{S}_k=\frac{\cos[2\pi/(2+k)]}{\cos[\pi/(2+k)]},~~~ ~~~(k\ge 2).
	\ee
	We emphasize that this result with a real scattering amplitude (being $\ge 0$ except for $k=1$ for which $\mathcal{S}_1=-1$) only holds at the charge degeneracy point, $\Delta E=0$.
	Plugging this result into Eq.~(\ref{eq:genericform}), in Fig.~\ref{fig:ABosc} we illustrate the conductance as a function of flux. 
	In a reference configuration with the QD detached from the first arm $(\gamma_1=0)$, $G_\Phi=r_1 r_2^* e^{-i \phi}+h.c.$. Namely, the visibility is given by $4|r_1 r_2|$ with an arbitrary phase dictated by details such as the interferometer length 
	[Fig.~\ref{fig:ABosc}(a)]. 
	
	\paragraph*{1CK effect.}  Connecting the QD only to the first arm, ($\gamma_1\ne 0$, $\gamma_2=\gamma_3=0$), the 1CK effect develops. This is an example of a FL fixed point described in terms of a scattering phase shift $\delta$, such that $\mathcal{S}=e^{2 i \delta}$ is  unitary. For the 1CK effect at the charge degeneracy point $\Delta E=0$, corresponding to zero Zeeman field in the usual ``spin" Kondo system, $\delta=\pi/2$, hence $\mathcal{S}=e^{2 i \delta}=-1$. This leads to a unit relative visibility and a $\pi$-shift of the interference term, as in Fig.~\ref{fig:ABosc}(b). 
	
	A finite gate detuning $\Delta E \ne 0$ leads to a deviation of the average charge of the island from $N+1/2$. As $\Delta E$ varies, the charge of the QD $N(\Delta E)$ displays Coulomb steps for small enough $\gamma_i$'s~\cite{furusaki1995theory}. In a FL at $T=0$ the phase shift is directly related to the charge via the Friedel sum rule, $\delta (\Delta E)= \pi N(\Delta E)$. In an experiment, probing the phase via the MZ interference, while simultaneously capacitively measuring on-site charge would test that this sum rule holds. As the QPCs become more and more open (i.e. upon increasing the $\gamma_i$'s), the charge as a function of gate voltage should gradually increase with a constant slope up to weak Coulomb oscillations, as can be observed using these two measurements.
	
	At $T=0$ the variation of the AB phase through the MZ interferometer upon varying the gate detuning $\Delta E$ can be accounted for by a noninteracting scattering model. However, the key property that cannot be explained by a single-electron scattering model is the reduction of the visibility occurring in multichannel case.

	\paragraph*{Symmetric MCK fixed points.} We proceed by coupling additional leads, but first setting the QD to charge degeneracy, $\Delta E=0$.  Upon coupling to the QD a second channel ($\gamma_1=\gamma_2 \ne 0$, $\gamma_3=0$), as in Fig.~\ref{fig:ABosc}(c), the visibility vanishes, $\mathcal{S}_{k=2}=0$, 
	and thus the interference term completely disappears as $T,V \to 0$. In other words, in contrast to the previously-considered realization of the 2CK with a single electron quantum dot~\cite{carmi2012transmission}, here the unitary paradox of the 2CK NFL state~\cite{maldacena1997majorana} directly manifests in the visibility.
	
	We note that the experimental results in  Duprez \emph{et. al.}~\cite{duprez2019transmitting} are reminiscent of the sequence of behaviors in Fig.~\ref{fig:ABosc}(a,b,c) although a different interpretation was given in a different regime with large $r_{1,2}$, uncontrolled $\gamma_{1,2}$ 
	and unknown energy splitting between pseudospin (charge) states.
	
	\begin{figure*}[]
		\includegraphics[width=2\columnwidth]{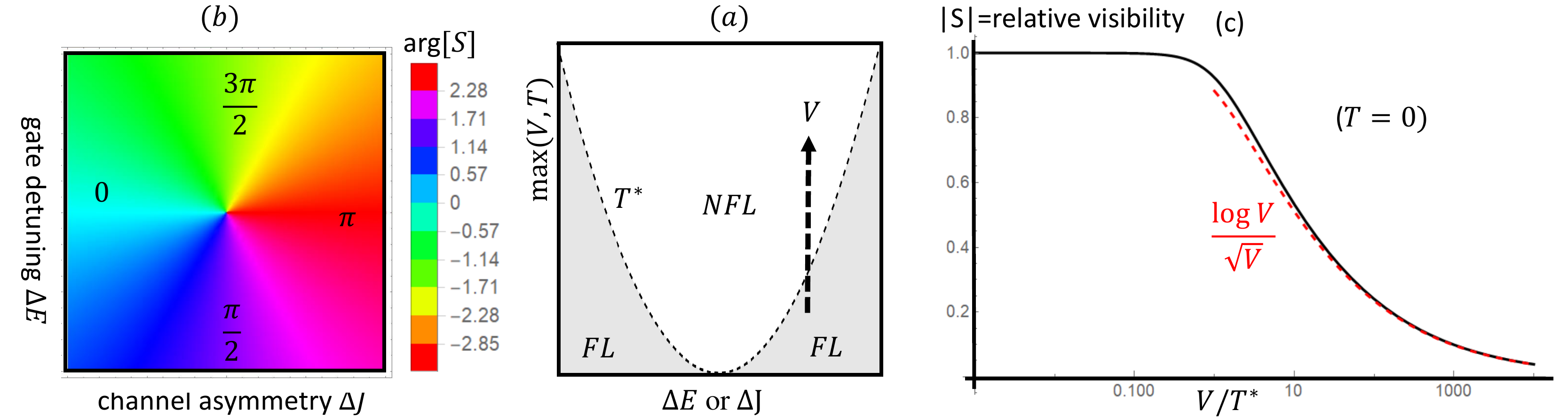}
		\caption{(a) 2CK quantum critical point, displaying an energy scale $T^*$ which vanishes at the critical point controlled by symmetry breaking perturbations $\Delta J$ and $\Delta E$. (b) The AB phase $\arg{\mathcal{S}}$ changes continuously, winding by $2\pi$, as the NFL state is encircled in the plane spanned by $\Delta J$ and $\Delta E$. (c) For $T^* \ll T_K$ the crossover of the visibility (as well as phase shift, not shown), can be computed exactly, see Eq.~(\ref{eq:exact}). }\label{fg:crossover}
	\end{figure*}
	
	Finally, upon coupling the QD to a third channel with equal magnitude ($\gamma_1=\gamma_2=\gamma_3 \ne 0$) as in Fig.~\ref{fig:ABosc}(d), the visibility approaches $\mathcal{S}_{k=3}=\frac{\sqrt{5}-1}{\sqrt{5}+1}=1-1/\varphi$ where $\varphi$ is the golden ratio, with the phase shift reverting to its reference value. Observing all the different behaviours shown in Fig.~\ref{fig:ABosc} in a single device, as couplings are tuned, could not be explained by a noninteracting single electron model.

	
	
	

	Below we discuss the effect of various symmetry breaking perturbations such as channel asymmetry or gate voltage detuning.

	\paragraph*{Symmetry breaking FL states.} We now consider the FL fixed points obtained by starting from the 2CK state $\gamma_1=\gamma_2$, $\gamma_3=0$, and either breaking the particle-hole symmetry by a gate voltage $\Delta E \ne 0$, corresponding to a Zeeman field on a conventional spin-based Kondo impurity, or by breaking the channel symmetry by a finite  $\Delta J=  \gamma_1 - \gamma_2 \ne 0$.
	Any  small deviation from the ``spin" and channel symmetries generates an energy scale $T^*$, such that upon lowering the temperature or voltage below $T^*$, the system undergoes a crossover from NFL to FL behavior~\cite{pustilnik2004quantum,mebrahtu2013observation,keller2015universal}, see Fig.~\ref{fg:crossover}(a). For the 2CK case the crossover scale is quadratic in the perturbations~\cite{pustilnik2004quantum,mitchell2012two,mitchell2016universality}
	\be
	\label{eq:TSTAR}
	T^*=\eta_1^2+\eta_2^2,~~\eta_1\propto 
	\Delta J,~~\eta_2 \propto
	\Delta E.
	\ee
	Manifestations of this FL-NFL crossover were studied experimentally both for spin-Kondo~\cite{keller2015universal} and charge-Kondo systems~\cite{iftikhar2015two}, as well as in other devices~\cite{mebrahtu2013observation}.

	

	We now consider the low energy limit, $T,V \to 0$. The phase shift in the FL regime  depends on the ratio between  channel anisotropy $\Delta J$ and gate voltage detuning $\Delta E$. For $\Delta E=0$ the FL  switches from a 1CK state occurring with channel 1 for $\gamma_1>\gamma_2$, in which case $\delta=\pi/2$, to one occurring with channel 2 for $\gamma_2>\gamma_1$, in which case $\delta=0$. These two values of the phase shift correspond to the $\Delta E=0$ line in Fig.~\ref{fg:crossover}(b).  It is interesting to explore the effect of encircling the NFL point $\Delta E=\Delta J=0$~\cite{sela2006adiabatic}. As seen in Fig.~\ref{fg:crossover}(b),  the phase shift performs a full winding by $\pi$ (with the definition $\mathcal{S}=e^{2 i \delta}$)~\cite{pustilnik2004quantum,sela2011exact,mitchell2012universal},
	\be
	\label{eq:phaseshift}
	\mathcal{S}_{FL}=-\frac{\eta_1+i\eta_2}{\sqrt{\eta_1^2+\eta_2^2}}.
	\ee
	In the pseudospin-polarized FL state for dominating $\Delta E$, the phase shift is $\pi/4$ or $3 \pi/4$ corresponding to a complex scattering amplitude $e^{2 i \delta}=\pm i$~\cite{pustilnik2004quantum}. Thus, the present MZ device can probe this phase shift winding around the NFL fixed point.
	
	\paragraph*{Exact NFL-FL crossover.}
	Based on the phase diagram in Fig.~\ref{fg:crossover}(a), one expects that the relative visibility of the 2CK MZ device will depend on the energy scale. Setting $T=0$ for simplicity, we expect a NFL region for $eV \gg T^*$ with vanishing visibility, and a FL region for $eV \ll T^*$ with unit relative
	visibility. Interestingly, 
	the 2CK model perturbed by any combination of channel asymmetry or Zeeman field maps to a known statistical mechanics problem known as the boundary Ising model~\cite{sela2012local} for which exact results are known~\cite{chatterjee1994local}. Borrowing these results, the Green function in the presence of a finite perturbation $T^*$ has been computed along each of these crossovers~\cite{sela2011exact,mitchell2012universal}. The $\mathcal{S}$-matrix is given by
	\be
	\label{eq:exact}
	\mathcal{S}=\mathcal{S}_{FL} \mathcal{G}(\omega/T^*),
	\ee
	where $\mathcal{G}(x)=\frac{2}{\pi} K[ix]$, $K[x]$ is the complete elliptic integral of the first kind, yielding asymptotically $\mathcal{G}(x)=1+ix/4-(3x/8)^2+\mathcal{O}(x^3)$ for $x \ll 1$; and $\mathcal{G}(x) =\frac{\sqrt{i}}{2 \pi} (\log [256 x^2]-i \pi)x^{-1/2}$ for $x \gg 1$.

	
	Plugging the exact Green function into Eq.~(\ref{eq:genericform}), we obtain
	the visibility and phase shift analytically. As shown in Fig.~\ref{fg:crossover}(c) the relative visibility given by $|\mathcal{G}(eV/T^*)|$ tends to 1 in the FL regime and decays as $1/\sqrt{V}$ for $eV \gg T^*$ in the NFL regime. Due to the complex structure of the function $ \mathcal{G}(\omega/T^*)$ the phase of the AB oscillations, $\arg (\mathcal{S}(eV))$ in Eq.~(\ref{eq:genericform}) also displays a crossover as a function of $V$ (not shown) as confirmed with NRG~\cite{mitchell2012universal}. 
	
	\paragraph*{Summary and outlook.} We analyzed an electronic Mach-Zehnder interferometer in the quantum Hall regime as a proposed setup to directly probe longstanding predictions on the scattering phase and
	sub-unitary scattering amplitude of single electrons in the multi-channel Kondo effect. The setup in general allows probing any quantum impurity model including multiple-impurity systems exhibiting exotic critical points as recently studied experimentally~\cite{pouse2021exotic,karki2022two}. An interesting future direction would be to use such an interferometer to probe the phase and possibly non-abelian statistics of Kondo anyons~\cite{lopes2020anyons,komijani2020isolating,PhysRevB.105.035151,lotem2022manipulating} by placing multiple dots along the chiral interference arm.
	
	\paragraph*{Acknowledgments.} ES, AA and FP acknowledge support from the European Research
	Council (ERC) under the European Unions Horizon
	2020 research and innovation programme under grant
	agreement No. 951541. ES acknowledges ARO (W911NF-20-1-0013) and the Israel Science Foundation, grant number 154/19. YO Acknowledges supportfrom the ERC
	under the European Union’s Horizon 2020 research and innovation programme
	(grant agreements LEGOTOP No. 788715), the DFG
	(CRC/Transregio 183, EI 519/7-1). DGG acknowledges support from U.S.Department  of  Energy,   Office  of  Science,   Basic  En-ergy  Sciences,  Materials  Sciences  and  Engineering  Di-vision,  under  contract  DE-AC02-76SF00515, at the initiation of the project, and from the Gordon and Betty Moore Foundation’s EPiQS Initiative through grant GBMF9460 at later stages. We thank Christophe Mora and Dor Gabai for discussions.

 \newpage
 
\widetext
 	\section{Supplementary information: Calculation of the current}
 	We write the tunneling Hamiltonian $H_T=T_1+T_2$, $T_1=\lambda_1 \psi^\dagger_0(-L)\psi_1(-L)e^{i V/t}e^{i \phi}+h.c.$, and $T_2=\lambda_2 \psi^\dagger_0(L)\psi_1(L)e^{i  V/t}+h.c.$.  We set $e=\hbar=v_F=1$. We obtain the current operator
 	\bea
 	I&=&\frac{dN_0}{dt}=i[H,N_0]=I_1+I_2,\nonumber \\
 	I_1&=&-i \lambda_1 \psi^\dagger_0(-L)\psi_1(-L)e^{i  V/t} e^{i \phi} +h.c. \nonumber \\
 	I_2&=&-i \lambda_2 \psi^\dagger_0(L)\psi_1(L) e^{i \phi} +h.c.
 	\eea
 	where $N_0=\int dx \psi^\dagger_0 \psi_0$ and $\{\psi_i(x), \psi_{i'}(x')\}=\delta_{ii'} \delta(x-x')$.
 	We evaluate the current
 	\be
 	\langle I(t=0) \rangle=-i \int_{- \infty}^0 dt \langle 0 | [I(0),H_T(t)] | 0 \rangle = I_0 + I_{\Phi},
 	\ee
 	where $I_0$ does not depend on $\phi$ and $I_{\Phi}$ contains terms  proportional to $e^{-i \phi}$. For $I_0$ one finds
 	\bea
 	I_0&=&-|\lambda_1|^2 \int_{-\infty}^\infty dt e^{i Vt} (\langle \psi^\dagger_0(-L) \psi_0(-L,t) \rangle  \langle \psi_1(-L) \psi^\dagger_1(-L,t) \rangle-\langle \psi_0(-L,t) \psi^\dagger_0(-L) \rangle \langle \psi^\dagger_1(-L,t) \psi_1(-L) \rangle) \nonumber \\
 	&+&\left(\lambda_1 \to \lambda_2,-L \to L \right).
 	\eea
 	For the free propagators we have
 	\bea
 	\langle \psi_i(x,t) \psi^\dagger_j \rangle=\langle  \psi^\dagger_j(x,t) \psi_i \rangle= \frac{\delta_{ij}}{2 \pi} \frac{1}{\delta+i(t-x)},
 	\eea
 	which gives
 	\bea
 	I_0=-(|\lambda_1|^2+|\lambda_2|^2) \int_{-\infty}^\infty dt e^{i Vt} \left( \frac{1}{(\delta-it)^2}- \frac{1}{(\delta+it)^2} \right)=\frac{V}{2 \pi}(|\lambda_1|^2+|\lambda_2|^2).
 	\eea
 	Using the Landauer formula, we associate reflection coefficients
 	\be
 	R_i = |\lambda_i|^2,~~~(i=1,2).
 	\ee
 	Reinserting units, this corresponds to a conductance of $\frac{I_0}{V}=\frac{e^2}{h}(|\lambda_1|^2+|\lambda_2|^2)$, and  $R_i = |\frac{\lambda_i}{\hbar v_F}|^2,~~~(i=1,2)$.
 	
 	For the interference term we have
 	\bea
 	I_\Phi= \lambda_1 \lambda_2^* e^{i \phi} \int_{-\infty}^\infty dt e^{i Vt} (\langle \psi^\dagger_0(-L) \psi_0(+L,t) \rangle  \langle \psi_1(-L) \psi^\dagger_1(+L,t) \rangle-\langle \psi_0(+L,t) \psi^\dagger_0(-L) \rangle \langle \psi^\dagger_1(+L,t) \psi_1(-L) \rangle) + h.c.
 	\eea
 	In the absence of a QD along the path between $x=-L$ to $x=L$ in channel 1, we obtain Eq.~(4) of the main text, 
 	\bea
 	\label{eq:genreral}
 	I_\Phi=- \lambda_1 \lambda_2^* e^{i \phi} e^{i V2L} \int_{-\infty}^\infty dt e^{i Vt} \left( \frac{1}{(\delta-i(t+2L))^2}- \frac{1}{(\delta+i(t+2L))^2} \right)=\frac{V}{2 \pi}(\lambda_1 \lambda_2^* e^{i \phi} +h.c.).
 	\eea
 	Eventually,
 	\be
 	I=I_0+I_{\Phi}=\frac{V}{2 \pi} |\lambda_1+\lambda_2 e^{i \phi}|^2.
 	\ee
 	
 	Eq.~(\ref{eq:genreral}) can be applied for the general case with a scattering of arm 1 through the QD. The frequency space, $G(\omega,-L,L)=G_0(\omega,-L,L)+G_0(\omega,-L,0) \mathcal{T}(\omega) G_0(\omega,0,L)$, translates simply to $G(\omega,-L,L)=G_0(\omega,-L,L) \mathcal{S}(\omega)$.  Eq.~(\ref{eq:genreral}) gives the current as the Fourier transform of the product of correlators. We see that the Fourier transform changes by the extra factor $\mathcal{S}(\omega)$ evaluated at $\omega=V$. It contains both an absolute value, dictating the visibility, and a phase.

\end{document}